\documentstyle[preprint,aps]{revtex}
\begin{document}
\draft
\title{Statistics of self-avoiding walks on randomly diluted lattice}
\author{M. D. Rintoul, Jangnyeol Moon\cite{present} and Hisao Nakanishi}
\address{Department of Physics, Purdue University, W. Lafayette, IN 47907}
\date{\today}
\maketitle
\begin{abstract}
A comprehensive numerical study of self-avoiding walks (SAW's) on
randomly diluted lattices in two and three dimensions is carried out.  The
critical exponents $\nu$ and $\chi$ are calculated for various
different occupation probabilities, disorder configuration ensembles, and
walk weighting schemes.  These results are analyzed and compared with those
previously available.  Various subtleties in the calculation and definition
of these exponents are discussed.  Precise numerical values are given
for these exponents in most cases, and many new properties are recognized
for them.

\end{abstract}
\section{Introduction}
In the past decade, the problem of the critical behavior of a polymer
\cite{Fl69,Ge79,De90} in a disordered medium has generated a great deal of
interest \cite{Ch81,De81,Kr81,Ha83,Ly84,Ro84,Ro87,Le88,Me89,Do91,Mo91,%
Na91,Va91,Va92,Gr93}. There are many different physical realizations
which can lead to such a problem.  These include polymers trapped
in a porous medium, gel electrophoresis  and size exclusion chromatography,
which deal with the transport of polymers through membranes with very small
pores.

One of the simplest theoretical models of this problem is that of a
self-avoiding walk (SAW) on a percolation cluster.  The SAW
incorporates many of the important characteristics of the polymer, such
as its flexible chain behavior and short distance repulsion, while the
percolation cluster represents the random medium.  Various analytical
methods have been used to attack this problem, such as mean field
theories \cite{Do91,Ob90,Ro90} and different types of renormalization group
calculations \cite{Ch81,Me89,Do91,Sa84}.  Unfortunately there has not
been a universal agreement among them, due to the difficulty in including
the geometrical effect of self-avoidance on a disordered medium.

Because of the difficulties in approaching the problem analytically,
there has been extensive computational work done in an attempt to get a
precise numerical estimate for many of the quantities which
characterize the SAW.  Despite the fact that the these numerical
calculations seem to represent a less complicated, more ``brute force''
approach to the problem, they too have many subtleties.  This has led
to a large volume of research with significantly different results for
seemingly very similar problems \cite{Le88,Me89,Mo91,Na91,Va92,Gr93,La90}.
In this paper, we present a comprehensive numerical study of this
controversial problem with a particular focus on the two critical exponents,
$\nu$ and $\chi$ (defined below).

A SAW is usually defined as a random walk which can never intersect
itself.  On a discrete lattice, the walk is constrained to move to a
nearest neighbor site during each time step, where the self-avoidance
condition further constrains the walk to occupy only sites which have
not been previously occupied.  The resulting random process now depends
in a complicated way on the history of the walk.  Because of this, the
formulas which govern the asymptotic behavior of the SAW's are
significantly different from those which govern those of free random
walks.

Written in terms of the quenched average, the asymptotic behavior for the
average mean-square end-to-end distance of a chain on a percolation cluster
becomes
\begin{equation}
\overline{\left< {R_N}^2 \right>} =
\frac{\sum_{\cal C} P({\cal C}) {\left< {R_N}^2 \right>}_{\cal C}}
{\sum_{\cal C} P({\cal C})} \sim N^{2 \nu}
\end{equation}
where the angular brackets indicate the average over all walks from
a given point on a given disorder configuration,
and the overbar indicates averaging over disorder.
For the latter average, ${\cal C}$ represents a disorder configuration,
and $P(\cal C)$ is the probability for that configuration to occur.
For the sake of specificity, we define ${\cal C}$ relative to
the fixed starting point of the SAW's.

Similarly, one also expects to be able to write the expression for
the average number of walks on the percolation cluster as
\begin{equation}
\overline{Z_N} = \frac{\sum_{\cal C} P({\cal C})
Z_N ({\cal C})}{\sum_{\cal C} P({\cal C})} \sim
N^{\gamma -1} {\mu}^N .
\end{equation}
For the full lattice ($p=1$), this form is known to be consistent with
renormalization group and other theoretical studies and with numerical
calculations of various types \cite{Ge79}. The value of $\gamma$ for the full
lattice is $43/32$ \cite{Ni82} in two dimensions and $\approx 7/6$ \cite{Ra85}
in three dimensions. Note, however, that for the case of $p < 1$,
$Z_N$ is a random variable with a multiplicative character
so that its mean and most probable values are far apart \cite{Mo91}.
This makes it intrinsically rather difficult to extract its true mean.
Moreover, under certain conditions on the distribution of $\ln Z_N$
\cite{Gi93}, this form itself may have to be replaced by another
asymptotic behavior which has a stretched exponential correction term
rather than the power law form $N^{\gamma -1}$.

The quenched average defined here is often referred to as
{\em one end fixed} \cite{Do91} even though the final configurational
average eventually accounts for all possible starting points; this is
because one end of the SAW's {\em is} fixed on a given disorder configuration
and the final disorder average would not be expected to change the
qualitative features of the average from any one {\em infinite}
disorder configuration. Such average may be related to a polymer
solution in a disordered space with one end grafted to (or trapped in)
a particular part of the substrate.  This situation should generally
lead to a stretched conformation since SAW's with the free end in
a dense region would dominate the average entropically where such regions
are further away from the fixed end than if density did not matter.

Another important exponent which is related to the disorder in the system
and has no full lattice analog is $\chi$, which is defined by the fluctuations
in the log of the number of walks as follows \cite{Do91}:
\begin{equation}
\mbox{var}[\ln Z_N ] = \overline{(\ln Z_N)^2} - (\overline{\ln Z_N})^2 \sim
N^{2 \chi}
\end{equation}
Recently, Le Doussal and Machta \cite{Do91} proposed the existence of a
new disorder fixed point for $p_c < p < 1$ based on renormalization
group arguments.  The existence of such a regime would be characterized
by a change in the value of $\chi$ for $p_c < p < 1$ from its value at
$p_c$ as well in the value of $\nu$ at $p_c$ from the corresponding values
at $p=1$ and $p_c < p < 1$.  Initial studies which were performed
to look for this crossover have not been fully consistent \cite{Do91,Va92}.

The aim of this work is to provide a comprehensive study of the two
critical exponents $\nu$ and $\chi$.
The behaviors of the mean square end-to-end distance and the variance
in $\ln Z_N$ are examined for the exponents $\nu$ and $\chi$ at
different occupation probabilities to obtain values both at the critical
percolation threshold and in the region where $p_c < p < 1$.  Averages over
all clusters and the {\em infinite} cluster are both calculated
in order to look for differences between the two.  A standard chain
averaging and a kinetically weighted average are both used to calculate
the exponents.  This allows a comparison between different physical
processes and also allows previous simulation results to be put in
proper perspective.  The exponent $\gamma$ was not
analyzed for in this work mainly because of the intrinsic difficulties
involved in extracting an average from a multiplicative random variable $Z_N$.
A full analysis of the distribution of $Z_N$ and the form of $\overline{Z_N}$
will be left for a future study.

The organization of this paper is as follows. In the next section, we
provide a brief summary of the numerical methods used in this work.
The enumeration data we have obtained are summarized and the exponent
estimates given in Section\ \ref{results} classified by the exponent
and by the type of weighting (see below).  In Section\ \ref{discussion}
we give detailed comparisons of our new estimates with the other
available results and discuss their significance in relation to the
previously controversial aspects of this problem.  Finally, we draw
some conclusions in Section\ \ref{conclusions}, clarifying the current
status of this problem.

\section{Numerical Method}
\label{methods}
In this section, we briefly describe the numerical approach we have used
in this work.  Our approach consists of two parts: we first randomly generated
the percolation disorder configurations by Monte Carlo simulation on grids
of prescribed sizes.  Then SAW's were generated by complete enumeration
on each disorder sample.  Below we give some details of these two operations.

While much of the computation was performed on standard sequential
computers, in those cases where the required CPU time per cluster
was small but the required number of clusters was very large,
we have resorted to massively parallel computation using a network of
$64$ Sun workstations.  This was done through a prototype parallel toolkit
called EcliPSe \cite{Re92}.  Some details of this aspect are given in
the Appendix.

\subsection{Generation of percolation clusters}
In this work, the percolation clusters were generated using site
percolation on the square and simple cubic lattice of edge length $L$.
Different values of $L$ were used for both two and three dimensions in
order to test finite size effects.  For the bulk of the data that did
not deal with finite size effects, the largest convenient value of $L$
was used.  In two dimensions, a value of $L=1000$ was normally used,
with the computer memory being the main constraining
factor.  In three dimensions, $L$ was taken to be $30$ mainly due to the
constraints imposed by the way we define the infinite cluster.

To construct a cluster, each site on the lattice was assigned to be
occupied with probability $p$.  As usual, occupied nearest neighbor
sites are connected together and the resulting connected components
of the lattice are called {\em clusters}.  Cluster connectivity was
determined using a breadth first search algorithm.  This was done
by going through the lattice and looking for sites that were not
already marked as belonging to a given cluster.  When such a site
was found, the entire cluster was {\em burned} \cite{He84} from that site
and a unique label was assigned to the elements in that cluster.

If an {\em infinite} cluster needs to be identified, the largest
cluster is searched for and checked to see whether or not it wraps
around the lattice in all directions when periodic boundary conditions
are applied.  If it does not, it is rejected and a new disordered
lattice is constructed.  Otherwise, it is accepted.  The wrapping
requirement was used for two reasons.  First, it is a commonly used way
to obtain a model for an infinite cluster on a finite system.  Also,
from a more practical standpoint, the wrapping condition allows us to
implement periodic boundary conditions in all directions and therefore
to choose the starting points for the walks without having to worry about
the lattice boundaries.  If the given simulation is concerned with all
clusters, then all of the clusters are used with periodic boundary
conditions.

\subsection{Generation of walks}
In the case of all cluster (AC) statistics, starting points for the
SAW's were chosen randomly from the entire lattice, with each cluster
being effectively weighted by the number of points in the cluster.  For
infinite cluster (IC) statistics, the points were chosen at random from
the {\em infinite} cluster.  Only one starting point was chosen per
lattice in this case.  This was done to prevent unexpected correlations
among the samples which would occur if different starting points were
too close to each other.  {\em All} walks less than or equal to a given
$N$ were then exactly enumerated.  The value of $N$ differed depending
on the other parameters of the calculation.  It was important to have
$N$ as large as possible in order to extract the asymptotic behavior,
but not so large that the enumerations would take an unreasonably large
amount of time to complete nor that the size of the SAW becomes
comparable to the lattice size.  Thus the values of $N$ ranged from
$25$ to $30$.

All walks from a given starting point were completely enumerated using
a very simple recursive procedure.  After an initial site was chosen, a
routine was called that marked that site as occupied and attempted to
walk to all unoccupied nearest neighbors.  If an unoccupied nearest
neighbor site was found, the same routine was again called with that
new site.  This happened until the walk had reached its maximum length
or could not find any available sites.  If this occurred, the current
site was marked as unoccupied and the walk retreated to its previous
step and continued to look for different available sites.  This method
was used to do all of the walk enumeration.

Two types of averaging were used.  The first type of averaging used was
{\em chain} averaging.  In this method, each walk from a given starting
point was given the same weight as every other walk of the same number of
steps from that same starting point.
This method is the one used in most enumerations and
best represents the model on which most theoretical calculations are
based.  In a polymer solution in random media, it would also represent
the case where the chain conformations are in equilibrium with the
environment.

The other type of averaging used was a kinetically weighted averaging
(previously called {\em walk} averaging \cite{Ma89}).  This
average weights the contribution of a given walk to a given measured
quantity (such as $\overline{\left<R^2 \right>}$) based on the inverse
of the connectivity of the sites in the walk.  Thus, the weight
$w_{\Gamma_N}$ for a given $N$-step walk $\Gamma_N$ can be written as
\begin{equation}
w_{\Gamma_N} = \frac{1}{z_0} \prod_{j \in \Gamma_N} \frac{1}{z_j - 1}
\end{equation}
where $z_j$ is the connectivity of site $j$ and the product is over all
sites except for the origin of the walk.  The factor of $z_j -1$ for $j
> 0$ reflects the exclusion of the site that the walk has just left.
The chain averaging is just equivalent to setting $w = C_N^{-1}$ for
all walks.  The {\em kinetic} average emulates the weighting of a standard
Monte
Carlo sampling method as applied to a diluted space\cite{Le88} .  It
may be described as a kinetic, growing walk (somewhat like the myopic
ant) except that it dies upon encountering its own track.  The kinetic
averages tend to have fewer fluctuations than the chain average for
similar samples.  In a polymer application, the kinetic averaging may
crudely correspond to the case where the chain is being grown {\em in
situ}.

Physically, the chain ensemble appears to be the appropriate one to
represent a dilute solution of polymer chains in porous media in
{\em equilibrium}.  However, the corresponding free walk problem showed
interesting differences between the walk and chain averages
\cite{Gi92}.  The investigation of possible differences for SAW's
appears warranted for this reason, as well as for a proper comparison
with some of the previous Monte Carlo results \cite{Le88}.

In order to extract the asymptotic exponent, the effective slope of the
plot of $\ln \overline{\left<R^2 \right> }$ vs. $\ln N$ was extracted
and plotted against $1/N$.  In the case of $\nu$, we have used
the effective exponent $\nu_N$ as given by the formula
\begin{equation}
2\nu_N = \frac{N \overline{\left<R_N^2 \right> }}
{\frac{1}{2}\left[\overline{\left<R_1^2 \right> }
+ \overline{\left<R_N^2 \right> }\right]
+ \sum_{i=2}^{N-1} \overline{\left<R_i^2 \right> }} - 1
\end{equation}
The effective exponent $\chi_N$ was extracted in a similar way, using
$\mbox{var}[\ln Z_N ]$.  Plotted in this way, the asymptotic value of
the exponent should be obtained as $N \rightarrow \infty$, or as $1/N
\rightarrow 0$.  The data is presented in this manner in order to
better observe the asymptotic behavior of the slope.  This also allows
us to estimate the error obtained in the extrapolation of the slope
to its $N=\infty$ value.

\section{Results of complete enumeration}
\label{results}
In this section, we describe the summaries of the data and the
exponent estimates obtained by analyzing them.  Detailed discussions on
their significance are presented in a separate section following the
current one.  The final exponent estimates are also given in
Tables\ \ref{table1} and \ref{table2} for the {\em chain} average and
in Tables\ \ref{table3} and \ref{table4} for the {\em kinetic}
average.

\subsection{Chain average results for $\nu$}
\label{chainnu}
In $d=2$, data were collected at $p=0.59273 \cong p_c$ for 6260
independent {\em infinite} clusters, which were divided into 3 different
runs.  These data are shown in Fig.\ \ref{fig1}.  The graph corresponds to
the effective slope plot that was described earlier.  This slope rises
gradually from 1.50 to 1.54 and then appears to level off to give an
asymptotic value of $\nu = 0.770 \pm 0.005$.  This general trend of
rising up and leveling off was seen in each of the three batches
separately.  The behavior in the case of the all cluster average also gave
an exponent estimate of $0.770 \pm 0.005$ and showed a similar increasing
behavior for larger $N$.  These data were taken from 14530 independent
disorder realizations.  The errors given
for the values of the exponents here and elsewhere were estimated by
using the scatter in the effective slope graphs for the overall
data as well as by considering the batch-to-batch fluctuations
in the effective slope graphs of the individual batches.

For the case of $p=0.65 > p_c$, the picture is similar to the $p=p_c$
case.  Because of the CPU time constraint involved in enumerating a
much larger number of walks than the $p=p_c$ case, we were forced to
use fewer starting points.  This was not a major setback since moving
away from the critical probability caused the fluctuations between
averages over different starting points to be smaller.  These data
(also shown in Fig.\ \ref{fig1}) seem to point to a value of $0.770 \pm
0.005$ for $\nu$.  The all cluster average is not expected to be
significantly different from the infinite cluster average for $p>p_c$
since the infinite cluster overwhelmingly dominates for $p > p_c$.

In the case of $d=3$ percolating clusters, we first took $p=0.3117 \cong
p_c$ and again performed calculations for both the {\em infinite} cluster
and for all clusters.  In both cases, the clusters were constructed on
a simple cubic lattice, where $L=30$.  The average was taken over 47830
independent realizations for the case of all clusters, and 20000
clusters for the infinite cluster case.  The resulting graphs for the
effective value of $\overline{\left<R^2\right>}$ are shown in
Fig.\ \ref{fig2}.  Both graphs are relatively straight, and the
asymptotic values of their slopes seem to differ.  A careful inspection
of the graphs show that $\nu = 0.660 \pm 0.005$ for the IC case, and
$\nu = 0.645 \pm 0.005$ for the AC case.  The values of the slopes seem
to differ more for larger $N$, which seems to imply that most of the
longer walks in the all cluster case are walks which are trapped in
isolated clusters (i.e., {\em not} in the {\em infinite} cluster).

For $p = 0.4 > p_c$, we also measured $\nu$.  This was done only on the
{\em infinite} cluster, but again, for $p$ this large the result should
not differ significantly from the all cluster result.  Again, the
disorder configuration was constructed on a  $L=30$ cubic lattice.
However, due to the increased density and the greater number of walks,
only 3920 clusters were used with $N \le 25$.  The graph of the
effective value of $2\nu$ is shown in Fig.\ \ref{fig2}.  The graph is
very straight, and shows an asymptotic value of $0.650 \pm 0.005$ for
$\nu$.

\subsection{Kinetic average results for $\nu$}
\label{kineticnu}
For all of the different types of disorder, the data obtained using
the kinetic average showed fewer statistical fluctuations in two
dimensions than their chain averaged counterparts.  The extrapolated
value for $\nu$ was slightly different between the two averages.  For
the IC case, we obtained an asymptotic value of $0.760 \pm 0.005$ for
$\nu$, slightly smaller than the chain average case.  The AC average
was almost identical to the IC case with a value of $0.755 \pm 0.005$,
which was again smaller than the chain average case.  Finally, the
$p=0.65$ case gave a value which was practically indistinguishable from
$3/4$.  The effective slope graphs which led to all three results are
shown in Fig.\ \ref{fig3}.

The case for d=3 is slightly harder to interpret, due to the steady
downward trend in the effective value of the exponent for the different
cases.  To extract the asymptotic value of the slope, we also looked at
the radius of gyration average for the same data.  In the asymptotic
limit, the average of the radius of gyration should have the same slope
as the mean square end-to-end distance.  The effective slope generated from
this data approached its asymptotic value from below and thus
gives an effective bound for the $N=\infty$ value (see also \cite{Mo91}).

Using the method described above, the value of $\nu$ on the infinite
percolating cluster was found to be $\nu = 0.645 \pm 0.005$. In the
case of all clusters at $p=p_c$, we calculated $\nu$ to be $0.625 \pm
0.005$.  For $p=0.4 > p_c$, the data also seemed to show an asymptotic
value of $\nu = 0.625 \pm 0.005$.  Like the two dimensional case, the
exponent data for the kinetic averages are all lower than their
respective chain averages.  These data are shown in Fig.\ \ref{fig4}.

\subsection{Chain average results for $\chi$}
\label{chainchi}
Our data for $\chi$ for $p=p_c$ and $d=2$ are taken from the same
calculations as for $\nu$ for the case of the {\em infinite} cluster.
These data are shown in Fig.\ \ref{fig5}.  For this case, our estimate
of $\chi$ is about $0.50 \pm 0.01$, which is consistent with a value
of  $1/2$.  In the case of the all cluster average, the data are taken
from a set that included 10000 disorder realizations with $N=25$ steps
for the SAW's.  It was necessary to take a larger sample of shorter
walks in order to reduce statistical fluctuations.  The data for all
clusters may well point to a slightly larger value for $\chi$.  Our
best estimate in this case is $\chi = 0.53 \pm 0.01$.  The data shown
in Fig.\ \ref{fig5} seem to strongly suggest that the IC and AC
averages are not the same.

The $p=0.65$ case in two dimensions in Fig.\ \ref{fig5} shows a
significantly different result than the $p=p_c$ case.  The data points
to a value of $0.47 \pm 0.01$ for $\chi$.  These data exhibit very little
statistical fluctuations and constitute a strong evidence for a new
disorder behavior for $p>p_c$, different from both that for $p_c$ or
the pure lattice behavior.

In three dimensions, we calculated $\chi$ using the same data sets as
were used in the calculation of $\nu$.  These data are shown in
Fig.\ \ref{fig6}.  The graph for the all cluster average is fairly
straight and yields a value of $\chi = 0.55 \pm 0.01$ in this case.
The exponent in the infinite cluster case is harder to extract since
there seems to be a downward trend in both the $p=p_c$ and $p=0.4$
case.  If these slopes are linearly extrapolated out to their
$N=\infty$ limit, one gets a value of $\chi = 0.49 \pm 0.01$ for the
$p=p_c$ case and $\chi = 0.27 \pm 0.02$ for the $p=0.4$ case.  We
believe, however, that this downward trend for large $N$ in the IC case
may be due to finite size effects for the case of $p=p_c$.  To show
this, we extracted the effective slope of $\chi$ for {\em infinite}
clusters built on simple cubic lattices of size $L=20$ and $L=25$ to
compare with the $L=30$ case.  All three cases are shown in
Fig.\ \ref{fig7}.  This graph explicitly shows the increasingly
downward behavior of $\chi$ for decreasing lattice size.

The downward trend in $\chi$ for $p = 0.4$ is much sharper, and
significantly different from the $p=p_c$ case.  In order to check
whether this was the asymptotic value for $\chi$ for all $p_c > p > 1$, we
also collected exact enumeration data for various $p$, where
$p_c \le p \le 0.4$.  These results (shown in Fig.\ \ref{fig8})
apparently indicate that the asymptotic value of $\chi$ decreases
with increasing occupation probability $p$.  This could even imply that
the asymptotic behavior of $\chi$ for $p>p_c$ does not follow a power law.

\subsection{Kinetic average results for $\chi$}
\label{kineticchi}
A corresponding definition of $\chi$ for kinetic averaging can be given by
\begin{equation}
\mbox{var}[\ln w_N] = \overline{(\ln w_N)^2} - (\overline{\ln w_N})^2 \sim
N^{2 \chi}
\end{equation}
where $w_N$ is the total remaining weight of all of the $N$ step walks
originating from a given starting point.  Thus, the {\em kinetic}
$\chi$ is defined by the variance of the log of the total amount of
weight associated with each walk.  Although the amount of weight starts
out with a value of 1, weight can be lost at each step if the walk is
{\em completely} trapped by the medium (and cannot grow) or if it
intersects itself.  This is different from the chain average where
weight is effectively lost from the chain average any time a vacant
site is encountered during the chain enumeration.

The data for $\chi$ at $p_c$ in two dimensions for the kinetic average
is shown in Fig.\ \ref{fig9}.  The IC average for $p=p_c$ can be
extrapolated to get a value of $0.48 \pm 0.02$.  Since the graph seems
to be leveling off, a value 1/2 is certainly not excluded.  Like the
chain average, the AC exponent is larger, but in the kinetic average
the difference between the IC and AC cases is much larger.  The AC
average points to an asymptotic value of $0.63 \pm 0.03$.  The
asymptotic value for $\chi$ when $p=0.65 > p_c$ seems to be falling off
very rapidly towards a value of $0.30 \pm 0.03$, but it is possible
that it could be continuing down even farther.

In three dimensions, the effective slope graph for the kinetically
averaged $\chi$ is even more similar to the chain averaged $\chi$, but
the difference between the IC and AC case is again more pronounced.
The AC case seems to be approaching a value of $0.66 \pm 0.02$, while
the IC value is approaching a value of $0.51 \pm 0.02$.  Like the chain
averaged $\chi$, the two and three dimensional results show qualitatively
very similar behavior for $p > p_c$.  The exponent $\chi$ for $p > p_c$
seems to be rapidly approaching a value of zero, which may suggest
a behavior that is not a power law for at least some values of $p>p_c$.
These data are shown in Fig.\ \ref{fig10}.

\section{Discussion}
\label{discussion}
In this section, we present a detailed comparison of our enumeration
results with other available estimates and discuss the new findings
in the broader perspective.  Unfortunately there seem to be no analytical
calculations which provide the relevant exponent estimates with any
reasonable level of accuracy.  Therefore, we first give detailed discussions
of the available numerical results, and at the end offer comparisons
with some theoretical predictions such as mean field and scaling arguments,
and certain renormalization group calculations.

\subsection{Exponent $\nu$ for the infinite cluster (IC) average at $p=p_c$}
Our present estimates of $\nu = 0.770 \pm 0.005$ in $d=2$ and $\nu =
0.660 \pm 0.005$ in $d=3$ for the chains on the infinite cluster at
$p=p_c$ represent a refinement of the values that were proposed in
several other works recently.  Our values are expected to be much more
precise than earlier estimates due to the fact that {\em all} walks
from a given starting point are enumerated, and the number of disorder
configurations chosen was very large.  This precision is best seen by
the fact that it passes the very stringent criterion of having the plot of
the effective slope as a function of the number of steps show very
little statistical fluctuation.  Vanderzande and Komoda\cite{Va92}
obtained the result $\nu = 0.77 \pm 0.01$ for chains on the IC using
the exact enumeration method.  They defined their infinite cluster as a
cluster which {\em spanned} the entire lattice, as opposed to our
condition that it {\em wrapped} in every dimension.  We tried both
types of requirements for our IC results and found little difference
between them.  The wrapping condition was chosen for our results since
it allowed us to use periodic boundary conditions and to choose our
starting point anywhere inside the cluster.  Infinite clusters defined
by the wrapping condition were also used in \cite{Mo91} for the
three dimensional case.  They obtained a value of $0.65 \pm 0.01$ for
$\nu$.

Many previous IC results were obtained using in whole or in part a
Monte Carlo simulation.  Among them, the work of Ref. \cite{Le88}
arrived at the conclusion that the asymptotic estimate of
$\nu$ changed very little by lattice dilution, even at the critical
dilution.  These simulations were performed on wrapping clusters as in
this work but with {\em kinetic} averaging.  Thus, we postpone the discussion
on those results toward the end of this subsection where we focus on the
kinetic averaging.

Some of the other previous works for IC configurations were performed
using a different type of {\em infinite} cluster.  In this situation
the infinite cluster was chosen based on a predetermined maximum
chemical distance from a seed site chosen on the cluster.  For some of
these enumerations, the clusters were grown with the standard Leath
algorithm \cite{Le76}. This algorithm starts with a seed site and grows
the cluster in shells around that site.  If any configuration that
contains at least one walk (no single isolated sites) is accepted, and
walks are enumerated from the seed, the resulting average is an AC
average.  However, if the configurations are only accepted after
growing a certain chemical distance $M$, then the clusters were
considered {\em infinite} clusters.  This method has an advantage in
that there are no boundary conditions to consider, and in some cases is
more computationally efficient than generating random numbers for the
occupation at each site of a large hypercubic lattice.  However, it does
have problems that the choice of $M$ which defines the IC is not as
directly related to the length scale of the cluster as L is when clusters
are generated on a periodic lattice.

This type of IC was recently used by Grassberger\cite{Gr93} to obtain
results for $\nu$ in two dimensions.  For this simulation, he chose only
those clusters with a chemical distance $> 200$.  This Monte Carlo
simulation was different than most previous ones because it first
evaluated the average connectivity of the cluster, and then chose the
sampling probability based on that.  For $\nu$ on the IC in two
dimensions, he calculated a value of $0.786 \pm 0.003$.   This number
was obtained by taking walks up to a maximum of 100 steps (and using
$O(10^4 )$ different configurations to average over disorder).  However,
even for $N$ this large, the average number of walks per sample was
approximately 5000.  Although the method Grassberger used to obtain
this value certainly should be sound, it seems unlikely that such a small
error bar could be assigned from a simulation that used so few walks
per disorder sample.

It is interesting to note that the kinetic average results on the infinite
cluster indicate a somewhat smaller $\nu$ than the chain averaged results.
The difference between the two cases is small, but appears to be
significant.  To understand why the kinetic average gives a smaller
exponent, it is important to note the main difference between the two
averages.  In the kinetic average, an $N$ step walk $\alpha$ will carry
with it a certain weight, $w_\alpha$, relative to the rest of the $N$
step walks.  All of the $N+M$ step walks which originate from $\alpha$
will have a total weight of $w_\alpha$, minus the weight of the walks
lost due to collisions and trapping.  In the chain average, however,
the relative weighting of all of the $N+M$ step walks will depend on
the total number of $N+M$ step walks.

Now consider a walk which leaves a very highly connected part of the
cluster, such as the backbone, and moves on to a dangling end.  This
walk, and all of the longer walks that it generates, will be trapped on
this dangling end, and won't be able to travel very far.  In the chain
average, these walks will not carry much of the weight since many more
walks will be generated on the highly connected part of the cluster.
However, in the kinetic average, these walks will still contribute
total weight equal to the weight of the original walk, until they
eventually begin to die out after getting trapped.  Because of this,
these trapped walks will contribute a much larger contribution to the
total average while they are alive.  Since the dangling ends occur on
all length scales, there will always be these contributions which keeps
the kinetic exponent lower.

A seemingly puzzling fact is that, for most values of $N$ considered in
this paper, the actual average end-to-end distance of the walks in the
kinetic average is larger than those from the chain averaging \cite{Mo91}.
This is because, for these values of $N$, the walks on the dangling ends
(which are more spread out and less compact than the rest of the  cluster)
contribute a larger end-to-end distance towards the average than
the chains on the dangling ends.  However, when these walks on
the dangling ends die out due to trapping, their contribution
towards the kinetic average is lost much faster than the contribution
to the chain average. The net result is a larger increase
in $\left<R^2\right>$ for the chains and therefore a larger $\nu$.
This, of course, means that, eventually for much larger $N$,
the value of $\left<R^2\right>$ must be larger for the chain averaging.

We tested this idea by performing the same two averages with the walks
constrained to move on the {\em backbone} of the infinite cluster.  The
backbone was defined as the multiply connected part of the {\em
infinite} cluster defined previously, when periodic boundary conditions
were applied.  Since the dangling ends are now removed, the walk and
chain averages should be approximately the same.  As seen in
Fig.\ \ref{fig11}, both graphs are almost exactly the same for large
values of $N$, and are both extrapolated to a value of $\nu = 0.775 \pm
0.005$.  This result is very similar to that of Woo and Lee
\cite{Wo91} which quoted a value of $0.77 \pm 0.01$ for $\nu$ on the
backbone of the {\em infinite} cluster on the square lattice.

We now return to the discussion of \cite{Le88}. These simulations
were carried out with statistical weights as in our {\em kinetic} averaging.
However, their results obviously do not agree with the present
calculation for the IC average at $p=p_c$.  A careful analysis of those data
in fact suggests that they were not representative of
the true average properties of the walk.  The main reason that the exponent
for the diluted case appeared to be the same as the full lattice is
believed to be as follows:  The specific Monte Carlo method that was
used attempted a fixed number of walks from each starting point.  For
large $N$, most of these walks died out on the diluted lattices.  If
all of the walks died out on a lattice for a given $N$, then it was not
counted in the ensemble average.  Unfortunately, there were so few
walks (on the order of several hundred) attempted from each starting
point that only the starting points which happened to be on the very
dense parts of the IC survived to contribute for large $N$.  In other
words, starting points in sparse regions were systematically excluded
from the averaging process.  Because of
this, the results were biased toward the dense clusters, and therefore
the $\nu$ that was measured was too small.

We have verified this scenario by performing simulations under the same
condition as in \cite{Le88}, and found that if more walks are
attempted from each starting point, a value of $\nu$ more consistent
with the enumeration results is obtained.  In fact, the number of
attempted walks from each starting point in \cite{Le88}
corresponds roughly to that for which on average only one walk would
survive the attrition at about the number of steps $N$ where those early
simulations reported a downward (or leveling) trend for their estimates
of the effective exponent $\nu_N$.  We also verified that having a large
number of starting points per cluster did not significantly affect the
results.

\subsection{Exponent $\nu$ for all cluster (AC) average at $p=p_c$}
The AC value of $\nu$ at $p=p_c$ has perhaps been the most studied and
most debated of all of the quantities discussed in this paper.  Our
estimate of $0.770 \pm 0.005$ in two dimensions shows that the value of
$\nu$ is definitely larger than the full lattice value of $0.75$.
Besides the data shown in Fig.\ \ref{fig1}, we also performed many
enumerations which used slightly smaller maximum lengths but many more
disorder configurations.  These runs also gave an asymptotic value of
$0.770$ for $\nu$ in the AC case, and showed no fluctuations that could
possibly point to a value at or lower than $0.75$.

There were also previous enumeration works \cite{Mo91} performed to
determine $\nu$ in two and three dimensions.  In those works,
the clusters were generated using the shell method described earlier,
requiring each disorder configuration to have at least 200 shells in two
dimensions.  Averaging over 1000 different disorder configurations, a value
of $\nu = 0.78 \pm 0.01$ was obtained.  Although the requirement that
the clusters have at least 200 shells causes the result to lie
somewhere between the AC and IC cases, this result is still a useful
one since both ensembles give approximately the same answer.  In
\cite{Gr93} Grassberger obtained a value of $0.785 \pm 0.003$ for $\nu$
in the AC, $p=p_c$ case, based on his incomplete enumeration.  While
our analysis does show that there could still possibly be some upward
trend in the estimates of $\nu$, we do not believe that its value would
be quite as large as in \cite{Gr93}.

These works seemed to settle the question of whether or not the value of
$\nu$ in two dimensions increases from the full lattice value to the
case of critical dilution.  However, there have been a few recent works
that still put the value for $\nu$ in the AC case to be possibly equal
to that of the fully occupied lattice.  In \cite{Va92}, Vanderzande and
Komoda estimated a value of $\nu = 0.745 \pm 0.010$.  Although this was an
exact enumeration performed in a way almost identical to our own, their
value was significantly lower.  This is possibly due to the fact that
they did not attempt to extract the asymptotic value of the slope.  For
smaller $N$ the value of the slope is less than $0.75$, so a
least-squares fit to the data that included these small $N$ values
could possibly yield a number that low.  Woo and Lee\cite{Wo91} also
gave a value of approximately 0.75 for the AC average.  This is probably
related to the fact that they used a Monte Carlo method with weighting more
similar to a kinetic average than to a chain average.  To test this, we
attempted a similar simulation with the same parameters as \cite{Wo91} and
also obtained a value of $\nu = 0.75$.

In three dimensions, the differences between the AC and IC averages at
$p=p_c$ are much more significant than the $d=2$ case.  As stated
earlier, this is probably due to the fact that the IC makes up a much
smaller fraction of the disorder configuration, and most walks find
themselves trapped on smaller isolated clusters.  Ref. \cite{Mo91}
obtained an estimate of $\nu = 0.65 \pm 0.01$ by an enumeration
of 5200 different clusters.  These clusters were grown with the
shell method and required to have at least 95 shells.  Vanderzande and
Komoda \cite{Va92} also studied the $d=3$ case with the enumeration method
and got a value of $0.635 \pm 0.010$ for the AC case, which is compatible
with our value within the error bars.  Unlike their result in two dimensions,
we would expect their estimate in three dimensions to be close to ours
even though they do not try to extrapolate their value out to $N = \infty$.
This is due to the fact that the effective exponent in the $d=2$ case has an
upward trend for larger $N$, while the effective exponent in the $d=3$ case
remains relatively constant for all $N$.

A different sort of exact enumeration was recently reported by Smailer
et al \cite{Sm93} to extract $\nu$ for a continuous disorder instead of
a discrete disorder as in percolation.  This means that the space is
not divided into regions of different, discrete statistical weights
(or equivalently {\em energy} cost) as in percolation (where accessible
sites have zero energy and the inaccessible sites have infinite energy
barrier), but each site or point in space is associated with a
continuously varying {\em energy} cost of hosting a part of the chain.
The basic assumption in this work appears to be that the finite
temperature asymptotic properties of the SAW in a quenched,
continuously random environment are governed by the zero temperature
fixed point, and that the properties of this latter fixed point would
correspond to the discrete percolation disorder at $p > p_c$.  They
then calculate the zero temperature properties in the continuous
disorder problem by looking at the minimum energy SAW by complete
enumeration.  More specifically, they considered the minimum energy SAW
of length $N$ (where $N \le 20$) in a square and simple cubic lattices where
the lattice sites were assigned energies based on a given distribution,
and then averaged the properties of the minimum energy walks over many
($\sim 10^5$) disorder configurations.  In two dimensions, two
different energy distributions were studied.  For a Gaussian
distribution with mean 0 and variance 1, a value of $\nu = 0.81 \pm
0.02$ was obtained, while a uniform distribution of energies between 0
and 1 gave a value of $\nu = 0.80 \pm 0.02$.  In three dimensions, just
the uniform distribution was studied and a value of $\nu = 0.71 \pm
0.03$ was calculated.

Although in their work \cite{Sm93} they compared the exponent estimates
they obtained with other, previous estimates for the percolation
disorder {\em at $p=p_c$}, they should be properly compared with the
percolation results for $p > p_c$. In any case, these values,
especially the value in three dimensions, seem to be much larger than
those obtained for the percolation disorder results for $p = p_c$
(this work as well as \cite{Mo91,Va92}),
and also more than those for $p>p_c$ from our present calculations (see
below).

\subsection{Exponent $\nu$ for $p>p_c$}
The result for $p = 0.65 > p_c$ in $d=2$ is also very interesting.  While
most previous works indicated that many of the asymptotic properties of
the SAW in the $p_c < p < 1$ regime should be the same as for $p=1$,
our data seem to indicate otherwise.  This was also true with the data from
\cite{Gr93}, although Grassberger again obtained a number which is
significantly larger than ours ($0.815 \pm 0.005$).  The data from
\cite{Va92}, however, gave a value of $0.75 \pm 0.01$ for this case. This is
somewhat lower than our value and happens to be equal to the value for
$p=1$.  We note that \cite{Va92} stated that
only a few hundred different starting points were used for large $p$.
Even though the fluctuation is generally less for $p>p_c$ and a smaller
sample space would give adequate results, for data in batches of that
size we certainly observed fluctuations large enough to account for the
difference.  In addition, this difference could also be partly due to
the fact that the slope was not extrapolated, as we discussed about
their data for the $p=p_c$ case.

Like the $d=2$ case, the value of $\nu$ for $p = 0.4 > p_c$ in three
dimensions was definitely not equal to the full lattice value, but was
found to equal $0.650 \pm 0.010$.  This is also very similar to the value of
$0.645 \pm 0.010$ reported in \cite{Va92}.

\subsection{Exponent $\chi$ for $p=p_c$ and $p>p_c$}
The values of $\chi$ obtained for the $p=p_c$ cases are important
because they represent a quantity which could be used to distinguish
between the IC and AC cases.  In both two and three dimensions, the AC
exponent is clearly larger than the IC exponent.  This is one of the
best instances of an exponent which gives a clear numerical difference
between these two cases.  Since the significance of $\chi$ was
mentioned only recently \cite{Do91}, there have been few numerical
studies done to measure it.  Vanderzande and Komoda calculated $\chi$ in
\cite{Va92}, but instead of using $\mbox{var}[\ln Z_N]$, they
used $\mbox{var}[\ln (Z_N +1)]$ \cite{VaPr}. Although the difference
between these quantities is small, it may be sufficient to affect the
estimate of $\chi$.  In fact, it is easy to show that the leading
order difference between these two terms is
\begin{equation}
2 \left<\ln(Z_N)\right> \left<\frac{1}{Z_N}\right> .
\end{equation}
This term is very small for large $Z_N$, but for small values of
$N$ it can be significant.  This point is illustrated in
Fig.\ \ref{fig12}.  Because of this, their data points for small $N$
are displaced downward and the fitted value of the slope is biased
toward a greater value.  Their estimates for the AC case of $\chi$ in
two and three dimensions at $p=p_c$ were $0.65 \pm 0.02$ and $0.64 \pm
0.02$, respectively, in contrast to our much smaller estimates (cf.
Table\ \ref{table2}).  When we used the same quantities,
$\mbox{var}[\ln (Z_N )]$ to test this idea, we obtained the graphs and
resulting exponent estimates quite similar to theirs.

Also, the behavior of $\chi$ for $p>p_c$ was examined.  The values
that we obtained for the specific values of $p$ ($p=0.65$ in $d=2$ and
$p=0.4$ in $d=3$) were much lower than for $p=p_c$, possibly
indicating a new disorder fixed point for $p_c < p < 1$.  Our estimates
appear to be consistent with those given in \cite{Va92}, (0.43 for $p=0.65$
in $d=2$ and 0.33 for $p=0.4$ in $d=3$) although no error bars were given in
their results.  We do not expect there to be a problem with the fact
that they were measuring $\ln (Z_N + 1)$ in this case since they
explicitly stated that they extracted $\chi$ from their data for larger
$N$.  Although it is very clear that $\chi$ indicates a different behavior
for $p>p_c$ than for $p=p_c$, it is {\em not} at all clear that there was a
similar behavior for all $p_c < p < 1$.  We note that for larger $p$,
it even seemed possible that $\mbox{var}[\ln Z_N ]$ did not follow
a power law behavior.

Finally we looked at $\chi$ for kinetically averaged walks in a
disordered medium.  This average describes walks that would be formed
if they were {\em grown} in the medium.  The behavior of $\chi$ for these
kinetically averaged walks displayed the same qualitative behavior as
those walks which were averaged according to the chain statistics.
However, the difference in $\chi$ at $p_c$ for the AC and IC cases was
much more pronounced in the kinetic case.

\subsection{Comparison with theoretical predictions}
So far, we have only discussed available numerical estimates in comparison
with our current results.  Here we will give a short discussion
of the comparison of our numerical results with the analytically obtained
estimates.  This discussion is only meant as a numerical comparison and
the reader is referred to the original works for the theoretical
arguments which led to the numbers we discuss.

In \cite{Do91}, Le Doussal and Machta used real space renormalization,
field theoretic renormalization, and Flory type mean field arguments
to study this problem.  As for the field theoretic studies,
they concluded that the previous study \cite{Me89} was incorrect
and that their own work was inconclusive
and more work was called for.  On the other hand, their real space
renormalization group study used hierarchical lattices to numerically
estimate the value of $\nu$ and $\chi$ on percolation clusters.
Their numerical estimates were significantly different than the ones
obtained here for percolation clusters on Euclidean lattices.
For $\nu$ (referred to as $\zeta$ in \cite{Do91}) at $p=p_c$, they
obtained a value of 0.850, and for $p_c < p < 1$, they find $\nu = 0.862$.
Although no direct claim was made as to what percolation system
their model corresponded to, the hierarchical lattice used was the one they
referred to as a {\em two dimensional} model.  However, there
seems to be no way to reconcile these numbers with the ones we
obtained and other numerical estimates available in the literature.  Even
their full lattice value of 0.847 does not seem to fit in with any usual
percolation system.  Their estimate of $\chi$ for that same system was
0.48 at $p=p_c$ and 0.29 for $p_c <p<1$, again much different from the
values we find (noting, particularly for $p_c$, that their results are
more closely interpreted as the AC ensemble results).
However, the prediction that the variation of $\nu$
between $p=1$ and $p_c < p < 1$ is about 2 \% is not too far from
what we find (2.7 \%), although we disagree with their prediction that
$\nu$ at $p_c$ is within 1 \% of its value for the full lattice.
In short, their real space renormalization results do not agree
with our present ones quantitatively although there are some qualitative
agreements.

The same paper also provides various mean field arguments to derive
expressions for $\nu$ and $\chi$ in terms of $d$ and $\nu_0$, the
pure lattice value for $\nu$.  These expressions were derived with no
consideration for the {\em critical} disorder, and thus should presumably
correspond to the case $p_c < p <1$ at best.

The first set of expressions they obtain are
\begin{eqnarray}
\nu_1(d) & = & \frac{1}{1 + d(1-\nu_o)/2} , \\
\chi_1(d) & = & {\nu_1}(1-d\nu_o/2) = \frac{1-d\nu_o/2}{1+d(1-\nu_o)/2} .
\end{eqnarray}
These expressions \cite{Do91} follow from the known scaling behavior
\cite{Ha87} of the probability distribution for the size of the pure SAW
in the {\em stretched} region and dimensional analysis.  The first formula
gives values of 0.8 and 0.618 for $\nu_1$ in two and three
dimensions, respectively.  These values are not very far from
those we obtained, especially in two dimensions.  However, the values
for $\chi_1$ is 0.2 in two dimensions and drops to 0.07 in three dimensions.
These values of $\chi_1$ do not correspond well to our estimates either
for $p_c < p <1$ or at $p=p_c$.

Their alternative Flory arguments led to several other possible
expressions.  For example, by expanding on the idea that the disorder
introduces an effectively attractive interaction among the {\em replicas}
(when the problem is formulated using the {\em replica} technique
\cite{St77}), they obtain
\begin{equation}
\chi_2 = \frac{2-d\nu_o}{3-d\nu_o} .
\end{equation}
This expression gives the values of $1/3$ for two dimensions and
0.19 for three dimensions. Again this prediction is quantitatively not
very good; however, the decreasing trend of $\chi$ for increasing dimension
(for $p_c < p <1$) does appear to be given correctly.  Yet another
Flory expression they provide is
\begin{eqnarray}
\nu_3 & = & \nu_o , \\
\chi_3 & = & 1 - d \nu_o /2  .
\end{eqnarray}
This set of expressions follow from a Flory argument where the effect of
random environment is treated only dimensionally.  The predicted values
are again not quantitatively acceptable for any value of $p$.

Obukhov \cite{Ob90} proposed a relation between $\chi$ and $\nu$,
\begin{equation}
\chi = 2 \nu - 1 .
\end{equation}
This result would follow if one assumed that the fluctuation $N^{\chi}$ in
$\ln Z_N$ were accounted for by the stretching entropy of the form
$R^2 /N$.  Like the relations proposed in \cite{Do91}, this expression
would apply for $p_c < p < 1$ and not for $p=p_c$, if at all.  Indeed,
our numerical values are in reasonably good agreement with this relation
for $p >p_c$ but not for $p=p_c$.
It is interesting to note that the goodness of this relation may suggest
that the effect of self-avoidance is only to rescale $R$ and the fundamental
relation between $R$ and the fluctuation in $\ln Z_N$ remains unaffected
(at least for $p_c < p < 1$).

While most of the above discussions in fact concerns $p>p_c$,
there exist also a large number of Flory type predictions for the
exponent $\nu$ at $p_c$ in the literature (but so far not for $\chi$).
Most of these arguments use the fractal dimension $d_f$ of the critical
percolation cluster or that of the backbone, $d_f^B$ to take account
of the underlying geometry.  This seems to imply that they are all
specifically for the IC ensemble and {\em not} for AC.  (Note, however,
that Ref. \cite{Ro87} discussed the relationship between the exponent
$\nu$ in the two ensembles and Ref. \cite{Ki90} which argued that they
are the same.)  Simplest of these mean field expressions is
\begin{equation}
\nu = \frac{3}{2+d_f} ,
\label{eq:kremer}
\end{equation}
which follows by substituting $d_f$ for the Euclidean dimension $d$
in the corresponding full lattice Flory expression \cite{Kr81}.
Somewhat ironically, this turns out to yield a very good numerical
agreement, giving 0.77 for two dimensions and 0.66 for three dimensions.

Many of the more complicated predictions, all of which arise from
some form of mean field approximation, were summarized in a brief paper by
Kim \cite{Ki92}. Roughly, there are two types of such approximations;
in one type, different formulas follow because of the different postulated
forms of the entropic contribution to the Flory free energy, while
in the other type, the argument is based on the dimer formation of
two SAW's. In both cases, the possibility of using either the full fractal
dimensionality $d_f$ or the backbone fractal dimensionality $d_f^B$
effectively doubles the number of different predictions.
There are far too many formulas to reproduce here and their predicted
values for $\nu$ at $p_c$ range from about 0.71 to 0.78 in two dimensions
and from 0.61 to 0.66 in three dimensions, the variation being partly
due to the uncertainties in the fractal dimensionalities needed to
evaluate the formulas.

The most notable feature of these Flory predictions is that they all
produce reasonable numerical predictions and, since there is no reason
in general to expect a Flory prediction to be {\em exact}, we cannot rule
out any of them simply because of a slightly worse agreement with the
numerical results.  This means that we cannot at this time rule in favor
of or against the respective assumptions in the form of the entropic term
in the free energy which resulted in different Flory formulas.  It is,
however, ironic that the most simple minded Flory approximation
Eq.\ (\ref{eq:kremer}) gives the best and almost exact agreement with
our results.  This might mean that the effect of critical disorder is
mainly to affect the scaling behavior of the interaction term and the
form of the entropic contribution largely remains unaffected.  Such a
scenario cannot be entirely correct, however, for it could not possibly
explain the results for $p>p_c$.

\section{Conclusions}
\label{conclusions}
Having performed a large scale complete enumeration work and
having made detailed comparisons with other available estimations
of the relevant exponents, we would like to offer some concluding remarks.

We believe that, this work, along with some other recent works,
definitely rules out the
possibility that the value of $\nu$ for critically disordered media is
unchanged from its full lattice value in two and three dimensions.  In
both cases, the value is distinctly larger.  We even find very strong
evidence that the value of $\nu$ is increased from the full lattice
value for $p$ significantly above the percolation threshold.  We also
find very strong evidence that the kinetic averaging gives a slightly
lower value of $\nu$ for some cases of disorder due to probability
trapping, and present numerical data to verify this explanation.
Also, the results of previous Monte Carlo simulation \cite{Le88}
which gave incorrect results are explained.

The exponent $\chi$ was also studied in great detail, and many
previously unknown properties were recognized.  The values of $\chi$ on
the infinite cluster and on all clusters are significantly different.
This is one of the clearest examples of a critical exponent giving
different results for these two ensembles.  Also, our initial numerical
studies of $\chi$ suggest that its behavior may not be the same for all
$p_c < p < 1$, and the variance of $\ln Z_N$ may not even follow a
power law in this regime.  However, definite conclusions on this point
cannot be drawn from the present work alone and must await further
investigation.

With regard to the status of theoretical understanding, it seems
appropriate to say that currently there exists {\em no} theory that is
quantitatively acceptable as the solution to this problem.  This would
include real space renormalization, field theoretic renormalization,
and various mean field approximations.
In particular, there seems to be no good mean field approximation for
the exponent $\chi$ either at $p_c$ or for $p_c < p <1$.  Many Flory
approximations for $\nu$ give reasonable predictions for $\nu$, but
the simplest possible, uncontrolled approximation happens to give the
best agreement.  We interpret this as an indication that the basic physics
of this problem is still not understood.

\acknowledgments
This work was supported in part by a grant from ONR.  We are grateful
to P. Grassberger, S. B. Lee, and J. Machta for
fruitful and stimulating discussions, and to C. Vanderzande for
clarifying communication.  In addition, some of the computations used
a parallel toolkit called EcliPSe developed by Rego, Sunderam and
Knop and were carried out with their critical assistance.  We are deeply
grateful for this help.

\newpage
\appendix
\section*{Massively parallel computation}
\label{parallel}
Many of the results described in this paper were obtained on a Kubota
Pacific Titan P-3000 mini-supercomputer.  The average CPU time taken
for an enumeration with a particular occupation probability and
disorder type was about 5000 minutes in total.  For some of the higher
occupation probabilities, a prototype KPC K-3400 Alpha AXP workstation
was used.  These platforms are of a standard sequential type.  However,
for some parts of the enumeration work, we have made use of massively
parallel computation over a distributed set of Sun workstations.

Specifically, a massively parallel algorithm was used to generate the data
for cases where the average run time per cluster was small, but the
number of clusters needed was very large.  This was done using a prototype
package called EcliPSe \cite{Re92}.  EcliPSe is a library which supports
concurrent execution of applications over machines which are connected
via a network as well as on parallel hardware such as the Intel iPSC860.
In our case, the enumeration task was spread over 64 Sun-4
workstations.  These machines were intended primarily for instructional
purposes and we were able to take advantage of the school vacation periods
when they were largely idle.

Parallelization of the algorithm was performed in collaboration with
the developers of the toolkit \cite{Kn93}, and was not difficult in
technical terms.  The task of generating a cluster, choosing a starting
point for the walks, and enumerating them was assigned to every machine.
As the machines finished their task and reported their results, they were
asked to repeat the process.  This was done until a preset number of clusters
were assigned.

In a practical sense, however, the notion of parallelizing a number of
processes whose run times vary greatly presents some problems.  This
was the case with the enumeration of the walks.  Most importantly, one
must choose a fixed number of walks and run the enumeration to
completion.  This can be understood by first looking at the case of
running the different batches sequentially.  If we stop the enumeration at
a random time and reject the enumeration from the current starting
point, we are more likely to be rejecting a longer run and are biasing
the statistics toward the clusters with fewer walks.  Similarly, if we
stop the simulation and accept the current starting point we are
biasing the results toward the clusters with more walks.

If we now consider a similar situation, but with a large number of
processors all working at once, the problem becomes worse as the
results from the shorter enumerations pile up very quickly.  If we
stop all the processes at a random time, and reject all of the
enumerations that all of the $N$ processors are working on, we are
effectively eliminating the $N$ longest enumerations from the results.
Since the distribution is so wide in the the case of SAW's on
disordered media, this makes the results of a prematurely terminated
job useless, unless there are many orders of magnitude more clusters
than processors.

Another practical problem which arose from the wide distribution of the
number of walks is that of parallelization efficiency.  Since some of
the enumerations took much longer than others, there was a great deal
of dead time at the end of the job as most of the processors were idle
while the last few were finishing up.  This was a significant problem since
the total number of starting points was only a couple of orders of magnitude
more than the number of processors.  In some cases, the amount of time
spent waiting for the last 10 enumerations to finish was as large as
the amount of time spent on all of the previous enumerations ($\approx$
5000).  This effect cut our parallelization efficiency from 64 (the
total number of processors) effectively to about 30 or 40, depending on
the parameters.  A way of reducing this effect would be to generate all of
the clusters and starting point {\em first}, and then estimating the
number of walks in each case using some sort of Monte Carlo method.
Then, the clusters could be ordered with respect to the estimated time
that they would take, and the larger ones could be started first.

\newpage
\begin{figure}
\caption{Effective slope plot of $\overline{\left< {R_N}^2 \right>}$ for
the {\em chain} averages on $d=2$ percolation clusters.  In the figure, the
symbols $\bigcirc$, $\triangle$, and $+$ denote $p=p_c$ for IC, $p=p_c$
for AC, and $p=0.65$ for IC, respectively.}
\label{fig1}
\end{figure}
\begin{figure}
\caption{Effective slope plot of $\overline{\left< {R_N}^2 \right>}$ for
the {\em chain} averages on $d=3$ percolation clusters.  In the figure, the
symbols $\bigcirc$, $\triangle$, and $+$ denote $p=p_c$ for IC, $p=p_c$
for AC, and $p=0.4$ for IC, respectively.}
\label{fig2}
\end{figure}
\begin{figure}
\caption{Effective slope plot of $\overline{\left< {R_N}^2 \right>}$ for
the {\em kinetic} averages on $d=2$ percolation clusters.  In the figure, the
symbols $\bigcirc$, $\triangle$, and $+$ denote $p=p_c$ for IC, $p=p_c$
for AC, and $p=0.65$ for IC, respectively.}
\label{fig3}
\end{figure}
\begin{figure}
\caption{Effective slope plot of $\overline{\left< {R_N}^2 \right>}$ for
the {\em kinetic} averages on $d=3$ percolation clusters.  In the figure, the
symbols $\bigcirc$, $\triangle$, and $+$ denote $p=p_c$ for IC, $p=p_c$
for AC, and $p=0.4$ for IC, respectively.}
\label{fig4}
\end{figure}

\begin{figure}
\caption{Effective slope plot of $\chi$ for
the {\em chain} averages on $d=2$ percolation clusters.  In the figure, the
symbols $\bigcirc$, $\triangle$, and $+$ denote $p=p_c$ for IC, $p=p_c$
for AC, and $p=0.65$ for IC, respectively.}
\label{fig5}
\end{figure}
\begin{figure}
\caption{Effective slope plot of $\chi$ for
the {\em chain} averages on $d=3$ percolation clusters.  In the figure, the
symbols $\bigcirc$, $\triangle$, and $+$ denote $p=p_c$ for IC, $p=p_c$
for AC, and $p=0.4$ for IC, respectively.}
\label{fig6}
\end{figure}

\begin{figure}
\caption{Effective slope plot of $\chi$ for {\em chain} averages on $d=3$
percolation clusters at $p=p_c$ on simple cubic lattices of various size.
Different symbols correspond to $L=30$ ($\bigcirc$), $L=25$ ($\triangle$)
and $L=20$ ($+$).}
\label{fig7}
\end{figure}

\begin{figure}
\caption{Effective slope plot of $\chi$ for AC {\em chain} averages on $d=3$
percolation clusters for various occupation probabilities $p$.  Different
symbols correspond to $p=0.3117$ ($\bigcirc$), $p=0.33$ ($\triangle$),
$p=0.35$ ($+$), $p=0.37$ ($\times$), and $p=0.4$ ($\diamond$).}
\label{fig8}
\end{figure}

\begin{figure}
\caption{Effective slope plot of $\chi$ for
the {\em kinetic} averages on $d=2$ percolation clusters.  In the figure, the
symbols $\bigcirc$, $\triangle$, and $+$ denote $p=p_c$ for IC, $p=p_c$
for AC, and $p=0.65$ for IC, respectively.}
\label{fig9}
\end{figure}
\begin{figure}
\caption{Effective slope plot of $\chi$ for
the {\em kinetic} averages on $d=3$ percolation clusters.  In the figure, the
symbols $\bigcirc$, $\triangle$, and $+$ denote $p=p_c$ for IC, $p=p_c$
for AC, and $p=0.4$ for IC, respectively.}
\label{fig10}
\end{figure}

\begin{figure}
\caption{Effective slope plot of $\overline{\left< {R_N}^2 \right>}$ for
{\em kinetic} and {\em chain} averages on the backbone of $d=2$ percolation
clusters at $p=p_c$.  In the figure, the symbols
$\bigcirc$ and $\triangle$ denote {\em chain} and {\em kinetic} averages,
respectively.}
\label{fig11}
\end{figure}

\begin{figure}
\caption{Plot of $\mbox{var}[\ln(Z^{eff}_N)]$ vs. $N$, where the symbol
$\bigcirc$ corresponds to $Z^{eff}_N \equiv Z_N$ (our method)
and the symbol $\triangle$
corresponds to $Z^{eff}_N \equiv Z_N + 1$ (as in \protect\cite{Va92}).}
\label{fig12}
\end{figure}

\newpage
\begin{table}
\caption{Estimates of critical exponents $\nu$ and $\chi$ for {\em chain}
averaging for the square lattice in two dimensions.  The data for $p_c$
were collected at $p=0.59273$.}
\label{table1}
\begin{tabular}{cccc}
$p$ & $p=p_c$ (IC) & $p=p_c$ (AC) & $p=0.65 > p_c$ (IC) \\ \hline
$\nu$ & $0.770 \pm 0.005$ & $0.770 \pm 0.005$ & $0.770 \pm 0.010$ \\
$\chi$ & $0.50 \pm 0.01$ & $0.53 \pm 0.01$ & $0.47 \pm 0.01$
\end{tabular}
\end{table}

\begin{table}
\caption{Estimates of $\nu$ and $\chi$ for the {\em chain} averaging
for the simple cubic lattice in three dimensions. The data for $p_c$
were collected at $p=0.3117$.}
\label{table2}
\begin{tabular}{cccc}
$p$ & $p=p_c$ (IC) & $p=p_c$ (AC) & $p=0.4 > p_c$ (IC) \\ \hline
$\nu$ & $0.660 \pm 0.005$ & $0.645 \pm 0.005$ & $0.650 \pm 0.005$ \\
$\chi$ & $0.49 \pm 0.01$ & $0.55 \pm 0.01$ & $0.27 \pm 0.01$
\end{tabular}
\end{table}

\begin{table}
\caption{Estimates of critical exponents $\nu$ and $\chi$ for {\em kinetic}
averaging for the square lattice in two dimensions.  The data for $p_c$
were collected at $p=0.59273$.}
\label{table3}
\begin{tabular}{cccc}
$p$ & $p=p_c$ (IC) & $p=p_c$ (AC) & $p=0.65 > p_c$ (IC) \\ \hline
$\nu$ & $0.760 \pm 0.005$ & $0.755 \pm 0.005$ & $0.750 \pm 0.005$ \\
$\chi$ & $0.48 \pm 0.02$ & $0.63 \pm 0.03$ & $0.30 \pm 0.03$
\end{tabular}
\end{table}

\begin{table}
\caption{Estimates of $\nu$ and $\chi$ for {\em kinetic} averaging
for the simple cubic lattice in three dimensions. The data for $p_c$
were collected at $p=0.3117$. $\chi$ for $p=0.4$ suggested a possibly
non-power law behavior.}
\label{table4}
\begin{tabular}{cccc}
$p$ & $p=p_c$ (IC) & $p=p_c$ (AC) & $p=0.4 > p_c$ (IC) \\ \hline
$\nu$ & $0.645 \pm 0.005$ & $0.625 \pm 0.005$ & $0.625 \pm 0.005$ \\
$\chi$ & $0.51 \pm 0.02$ &$0.66 \pm 0.02$ & $-$
\end{tabular}
\end{table}

\end{document}